\documentclass[10pt,twocolumn,letterpaper]{article}

\usepackage{cvpr}              % To produce the CAMERA-READY version
\usepackage[accsupp]{axessibility} % Improves PDF readability for those with visual impairments.
\usepackage[dvipsnames]{xcolor}

\definecolor{cvprblue}{rgb}{0.21,0.49,0.74}
\usepackage[pagebackref,breaklinks,colorlinks,citecolor=cvprblue]{hyperref}
\usepackage{listings}
\def\paperID{8340} % *** Enter the Paper ID here
\def\confName{CVPR}
\def\confYear{2024}

\title{SonicVisionLM: Playing Sound with Vision Language Models}

\author{Zhifeng Xie$^{1,2}$, Shengye Yu$^{1}$, Qile He$^{1}$, Mengtian Li$^{1,2}$\footnotemark[2]\\
$^{1}$Shanghai University\\
$^{2}$Shanghai Engineering Research Center of Motion Picture Special Effects\\
{\tt\small \{zhifeng\_xie,yussisy,shu\_hql,mtli\}@shu.edu.cn}}
\begin{document}
\maketitle
\renewcommand{\thefootnote}{\fnsymbol{footnote}}
\footnotetext[2]{Corresponding author.}
\begin{abstract}
There has been a growing interest in the task of generating sound for silent videos, primarily because of its practicality in streamlining video post-production.
However, existing methods for video-sound generation attempt to directly create sound from visual representations, which can be challenging due to the difficulty of aligning visual representations with audio representations.
In this paper, we present \textbf{SonicVisionLM}, a novel framework aimed at generating a wide range of sound effects by leveraging vision-language models(VLMs). 
Instead of generating audio directly from video, we use the capabilities of powerful VLMs.
When provided with a silent video, our approach first identifies events within the video using a VLM to suggest possible sounds that match the video content.
This shift in approach transforms the challenging task of aligning image and audio into more well-studied sub-problems of aligning image-to-text and text-to-audio through the popular diffusion models.
To improve the quality of audio recommendations with LLMs, we have collected an extensive dataset that maps text descriptions to specific sound effects and developed a time-controlled audio adapter.
Our approach surpasses current state-of-the-art methods for converting video to audio, enhancing synchronization with the visuals, and improving alignment between audio and video components.
Project page: \url{https://yusiissy.github.io/SonicVisionLM.github.io/}
\end{abstract} 
\begin{figure}
  \centering
  \includegraphics[width=0.48\textwidth]{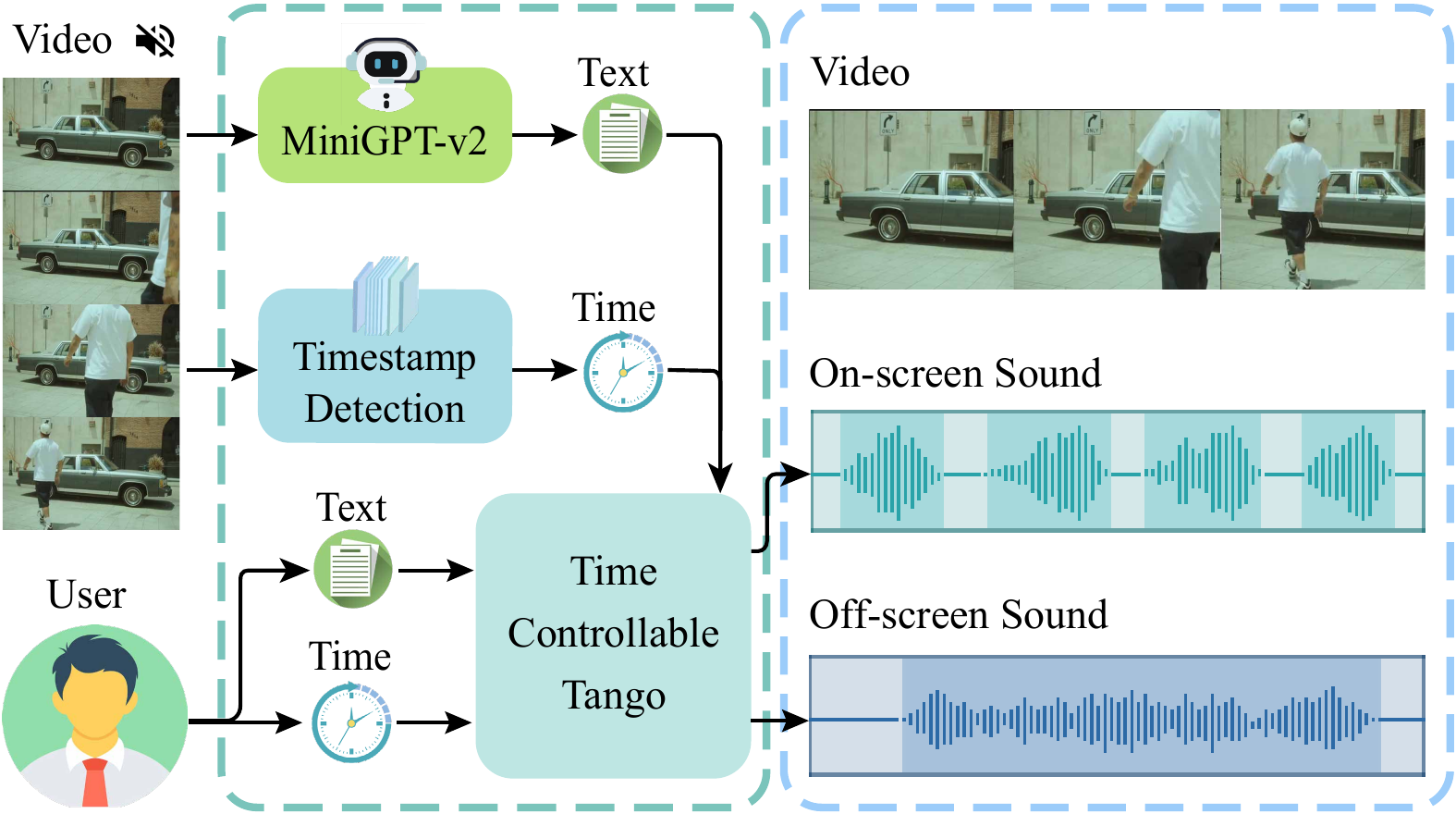}
  \setlength{\abovecaptionskip}{-2mm}
  \caption{A model implements the automatic detection of on-screen sound generation and accepts the user's editing of text and time in the off-screen section. On-screen sound refers to audio that originates from visible actions within the video frame. Off-screen sound is not directly observable on the screen.}
  \vspace{-6mm}
  \label{fig:teaser}
\end{figure}
\begin{figure*}
   \centering
   \includegraphics[width=1\textwidth]{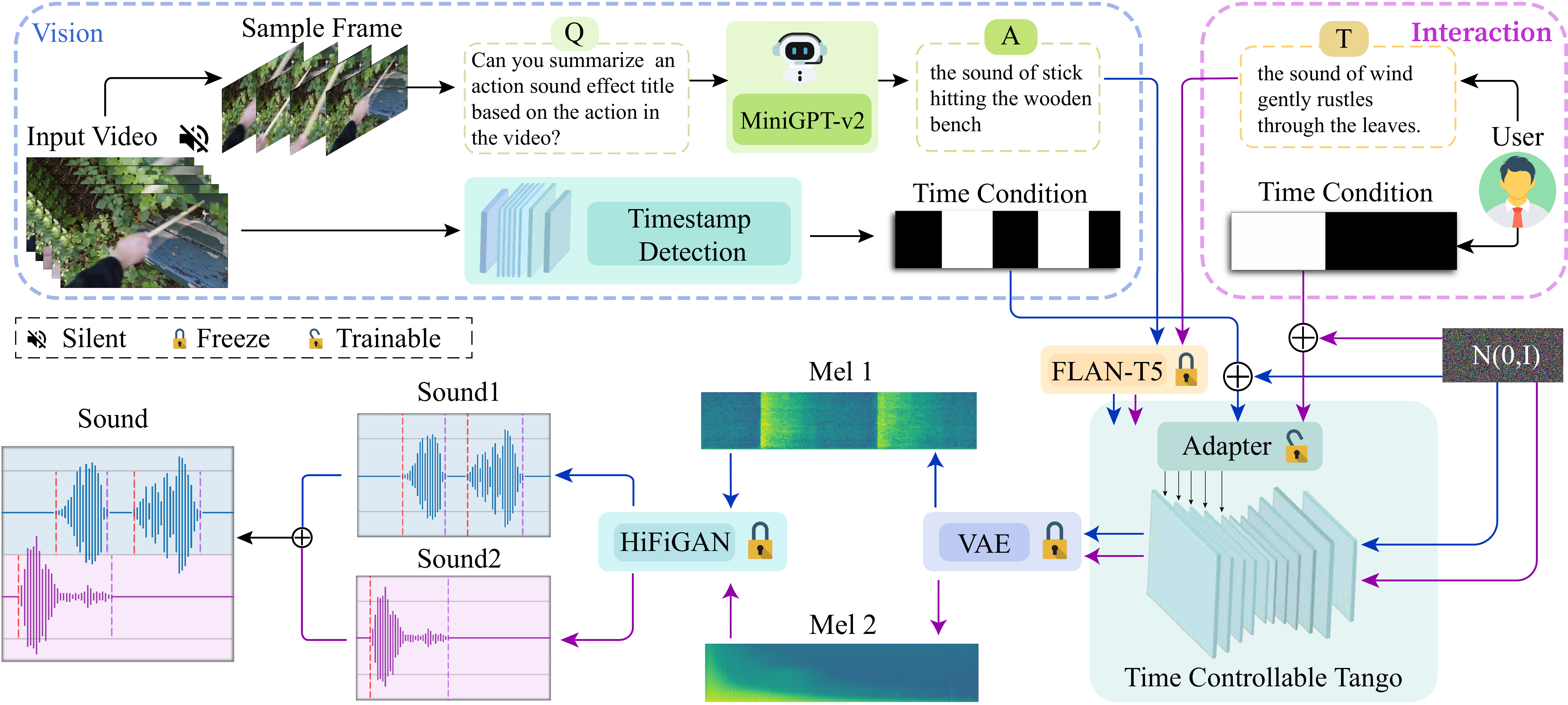}
   \setlength{\abovecaptionskip}{-2mm}
   \caption{\textbf{SonicVisionLM's framework.} SonicVisionLM presents a composite framework designed to automatically recognize on-screen sounds coupled with a user-interactive module for editing off-screen sounds. The blue dashed box and arrows in the figure represent the visual automation workflow: First, a silent video goes into the system to determine the occurring events (text) and their timing (time). Then, this information conditions the generation of sounds matching the screen. The purple dotted box and arrows show how users can modify or add off-screen sounds. }
    \vspace{-6mm}
   \label{fig:framework}
 \end{figure*}
\vspace{-2mm}
\section{Introduction}\label{sec:intro}
\vspace{-2mm}

\hspace{1em}The sound effects artists work with various types of sounds, including those visible on-screen (like footsteps or a car passing) and those audible but not visible (like background noises and heartbeats can enhance the video's authenticity and narrative).
On-screen sounds match what is happening in the video, while off-screen sounds establish the ambience and provide additional information.
Creating soundtracks for videos is a vital aspect of video production, but it can be labour-intensive for artists.
Therefore, the video-sound generation task has gained notable attention.

Although the recent approaches have made great efforts, the video-sound generation task is still challenging.
For on-screen sounds, achieving semantic relevance and maintaining temporal synchronization continues to be a complex issue.
It is hard to edit off-screen sounds.
Current methods \cite{shefferHearYourTrue2023,iashinTamingVisuallyGuided2021,luoDiffFoleySynchronizedVideotoAudio2023} primarily focus on the visual content to generate the corresponding sound, a subset of these methods \cite{duConditionalGenerationAudio2023a,cuiVarietySoundTimbreControllableVideo2023} considers editability.
Nonetheless, the alignment between video and audio features is tricky, leading to deficiencies including 1) incorrect sound meanings and mismatched timing, 2) monotonous sound effects, and a lack of complex scenarios.
Both lead to unsatisfactory results in the video-sound generation task.
 
We propose a novel framework named SonicVisionLM to solve the above deficiency, as shown in Fig.~\ref{fig:teaser}.
SonicVisionLM is proposed by introducing three key components: video-to-text, text-based interaction, and text-to-audio generation.
First, the video-to-text component focuses on generating sound effects for on-screen events.
This step uses a VLM to identify appropriate sound descriptions from the input silent video. 
Following this, a timestamp detection network is trained to extract specific temporal information from the video. 
A key innovation within this framework is the design of a time-conditioned embedding, which is utilized to guide an audio adapter. 
After that, the text-based interaction component allows users to change the text and timestamps from a previous video-to-text component or to input new corresponding text-timestamp pairs for personalized sound design.
Finally, the text-to-audio generation component accepts the text and timestamp conditions and inputs them into the LDM and adapter to generate diverse, time-synchronized, and controllable sounds.
Simultaneously, We collect a text-to-single-sound dataset, named CondPromptBank, for sound effects caption and timing cues, comprising over ten thousand data points, covering 23 categories.
The main contributions of this work are:
 \begin{itemize}
    \item We propose a novel framework called SonicVisionLM and collect a dataset CondPromptBank specifically for training a time-controllable adapter.
    It ensures the generated sound aligns perfectly with our text input and maintains precise timing control.
    \item We introduce three pioneering components: video-to-text, text-based interaction, and text-to-audio generation.
    This unique combination facilitates the automatic recognition of on-screen sounds while enabling user customization of off-screen sounds.
    \item The proposed framework achieves state-of-the-art results on conditional and unconditional video-sound generation tasks.
    The conditional task can be noticeably enhanced in all metrics. (IoU: 22.4\(\rightarrow \)39.7)
 \end{itemize}

\section{Related Work}
{\bf Audio Generation} can be broadly classified into two categories.
The \underline{\textit{Text-to-Audio Generation}} field includes Text-to-Speech (TTS) and Text-to-Music (TTM). 
Leading TTS models, such as FastSpeech2 \cite{renFastSpeechFastHighQuality2022} and NaturalSpeech \cite{NaturalSpeechEndtoEndText2022}, now produce speech virtually indistinguishable from human speech.
In TTM, MusicLM \cite{agostinelliMusicLMGeneratingMusic2023}, Noise2Music \cite{huangNoise2MusicTextconditionedMusic2023}, MusicGen \cite{copetSimpleControllableMusic2023a} and MeLoDy \cite{lamEfficientNeuralMusic2023b} are aimed to generate music segments from text, bringing innovation to music composition and synthesis.
Models like AudioGen \cite{kreukAudioGen}, AudioLDM \cite{liuAudioLDMTexttoAudioGeneration2023}, Tango \cite{ghosalTexttoAudioGenerationUsing2023}, and Make-an-Audio \cite{huang2023make} focuses on universal audio generation modeling.
AudioGen \cite{kreukAudioGen} treats audio generation as a conditional language modelling task, while the other three models employ latent diffusion methods to accomplish sound generation.
Current methods use datasets including sound effects, voices, and music, but practical applications use these elements separately. As the textual descriptions of time are subjective, videos are more intuitive and precise, so V2A requires more accurate semantic features and time control than T2A.
Therefore, we have created a text-to-single-sound dataset called CondPromptBank with detailed semantic segmentation and temporal annotations that help models produce high-quality sound effects for videos.
In \underline{\textit{Video-to-Audio Generation}} task, SpecVQGAN \cite{iashinTamingVisuallyGuided2021} utilizes a Transformer-based autoregressive model, drawing on ResNet50 or RGB+Flow features to generate sound.
Im2Wav \cite{shefferHearYourTrue2023} uses a dual-transformer model conditioned on CLIP features for sound generation.
CondFoleyGen \cite{duConditionalGenerationAudio2023a} and VARIETYSOUND \cite{cuiVarietySoundTimbreControllableVideo2023} introduce tasks for controllable timbre generation.
Diff-foley \cite{luoDiffFoleySynchronizedVideotoAudio2023} uses contrastive audio-visual pretraining to align audio and visual features.
ClipSonic \cite{dong2023clipsonic} learns the text-audio correspondence by leveraging the audio-visual correspondences in videos and the multi-modal representation learned by pre-trained VLMs.
However, the sounds generated by these methods often suffer from poor audio-visual synchronization, noticeable noises, and lack of editability.
Unlike the works above, our model ensures audio-visual synchronization, enriches diversity, and supports personalized user edits.
Our model provides a more comprehensive sound solution for video production.

\noindent{\bf Diffusion Model} has been utilized for generating both mel-spectrogram generation \cite{Grad-TTS,chen2022resgrad}, and waveform generation \cite{lamBilateralDenoisingDiffusion2021,leePriorGradImprovingConditional2022,Infergrad}.
However, their iterative generation process can be slow for high-dimensional data. 
Models such as AudioLDM \cite{liuAudioLDMTexttoAudioGeneration2023}, Make-An-Audio \cite{huang2023make}, and Tango \cite{ghosalTexttoAudioGenerationUsing2023} have successfully trained diffusion models within a continuous latent space.
Nevertheless, achieving satisfactory results in controlling LDM for audio generation tasks remains challenging.
This paper aims to introduce time control to ensure audio-visual synchronization.

\noindent{\bf Vision Language Models} like ChatGPT-4, which demonstrated advanced multi-modal abilities and inspired vision-language LLMs.
Vision-LLM \cite{wang2023visionllm} and LLaVA \cite{touvron2023llama} focus on aligning image inputs with large language models
Vicuna\cite{chiang2023vicuna} exhibit similar multi-modal capabilities.
Recent developments in this field include MiniGPT-v2 \cite{chenMiniGPTv2LargeLanguage2023}.
Kosmos-2 \cite{peng2023kosmos} demonstrates multi-modal LLMs' ability to perform visual grounding.
In this paper, we first introduce the VLMs to the audio generation task.

\section{Method}
\subsection{Overview}
\hspace{1em}In this section, we introduce the framework of SonicVisionLM, as shown in Fig.~\ref{fig:framework}.
Before delving into the specific design details, we first briefly overview the preliminary knowledge (Sec.~\ref{subsec:Preliminaries}). 
Then, we introduce the Visual-to-Audio Event Detection Module (Sec.~\ref{subsec:text}), which obtains textual descriptions of on-screen sounds through VLMs. 
Subsequently, we present the Sound Event Timestamp Detection Module (Sec.~\ref{subsec:time}), designed to accurately detect the timing information through network architecture.
Finally, we introduce the proposed time-controllable adapter as an extension of the audio diffusion model (Sec.~\ref{subsec:ldm}), enabling the generation of multiple sounds that are semantically coherent and temporally aligned.
\subsection{Preliminaries}
\label{subsec:Preliminaries}
{\bf Audio Diffusion Model.}
The text-prompt encoder encodes the input description $ \tau\in R^{L \times d_t}$ of the sound, where $L$ is the token count and $d_t$  and is the token-embedding size.
The latent diffusion model (LDM) is used to construct the audio prior $z_0 $  with the guidance of text encoding $\tau$. This essentially reduces to approximating the true prior $q\left(z_0\mid \tau \right)$ with parameterized $p\left(z_0\mid \tau \right)$.
LDM can achieve the above through forward and reverse diffusion processes.
The forward diffusion is a Markov chain of Gaussian distributions with scheduled noise parameters $0 < \beta_1 < \beta_2 < \cdot \cdot \cdot < \beta_N < 1 $ to sample noisier versions of $z_0$, where $N$ is the number of forward diffusion steps. For each step $n$, we define $\alpha_n = 1-\beta_n$, and calculate the cumulative product $\bar{\alpha}_n=\prod_{i=1}^n \alpha_n$.The diffusion equations are described as follows:
\vspace{-2mm}
\begin{equation}
\begin{aligned}
q\left(z_n \mid z_{n-1}\right) & =\mathcal{N}\left(\sqrt{1-\beta_n} z_{n-1}, \beta_n \mathbf{I}\right),
\end{aligned}
\label{eq:equation1}
\end{equation}
\vspace{-4mm}
\begin{equation}
\begin{aligned}
q\left(z_n \mid z_0\right) & =\mathcal{N}\left(\sqrt{\bar{\alpha}_n} z_0,\left(1-\bar{\alpha}_n\right) \mathbf{I}\right),
\end{aligned}
\label{eq:equation2}
\end{equation}
where the noise term $\epsilon$ and the final step of the forward process yields $z_N$ follow a Gaussian distribution, specifically $\epsilon, z_N \sim \mathcal{N}(0, I)$.
The reverse process denoises and reconstructs $z_0$ through text-guided noise estimation $( \hat{\epsilon}_{\theta})$ using following loss function:
\vspace{-2mm}
\begin{equation}
\mathcal{L}_{D M}=\sum_{n=1}^N \gamma_n \mathbb{E}_{\epsilon_n \sim \mathcal{N}(\mathbf{0}, \mathbf{I}), z_0}\left\|\epsilon_n-\hat{\epsilon}_\theta^{(n)}\left(z_n, \tau\right)\right\|_2^2,
\label{eq:equation4}
\end{equation}
After training LDM, we generate audio latent by sampling through the reverse process with $z_N \sim \mathcal{N}(0, I)$, conditioned on the given textual representation $ \tau $.
Its reverse dynamics are shown below:
\vspace{-3mm}
\begin{equation}
\begin{aligned}
p_\theta\left(z_{n-1} \mid z_n, \tau\right) =\mathcal{N}\left(\mu_\theta^{(n)}\left(z_n, \tau\right), \tilde{\beta}^{(n)}\right),
\end{aligned}
\label{eq:equation6}
\end{equation}
\vspace{-4mm}
\begin{equation}
\mu_\theta^{(n)}\left(z_n, \tau\right) =\frac{1}{\sqrt{\alpha_n}}\left[z_n-\frac{1-\alpha_n}{\sqrt{1-\bar{\alpha}_n}} \hat{\epsilon}_\theta^{(n)}\left(z_n, \tau\right)\right],
\label{eq:equation7}
\end{equation}
\vspace{-2mm}
\begin{equation}
\begin{aligned}
\tilde{\beta}^{(n)} & =\frac{1-\bar{\alpha}_{n-1}}{1-\bar{\alpha}_n} \beta_n .
\end{aligned}
\label{eq:equation8}
\end{equation}

The noise estimation $( \hat{\epsilon}_{\theta})$  is parameterized with U-Net framework \cite{ronneberger2015unet} with a cross-attention component to include the text guidance $\tau$. 

After that, the decoder of audio VAE \cite{kingma2022autoencoding} constructs a mel-spectrogram feature from the latent audio representation $ \hat{z}_0 $. 
This mel-spectrogram feature is conveyed to a vocoder to generate the final audio. 
\vspace{-1mm}
\subsection{Visual-to-Audio Event Understanding Module}
\vspace{-1mm}
\label{subsec:text}
\hspace{1em}The diversity of sound is influenced by various factors such as its source, actions, the environment, and more.
At the same time, these factors are often included in the description of visual images.
Inspired by the widespread use of VLMs \cite{liVideoChatChatCentricVideo2023,wangInternVidLargescaleVideoText2023,zhangVideoLLaMAInstructiontunedAudioVisual2023}, we chose MiniGPT-v2 \cite{chenMiniGPTv2LargeLanguage2023} to process visual information and generate descriptions of sounds.
Recognizing that MiniGPT-v2 was initially developed for single-image understanding and its limitations in conveying dynamic information, we have adapted the LLaMA-2 \cite{touvron2023llama} conversation template design to suit a multi-modal instructional framework.
To not miss sound events, we encoded four random video frames with time-sensitive prompts (temporal cues):
%
% The adapted template is defined as follows:
\begin{lstlisting}
First     ,< Img >< ImageFeature >< Img >.
Then      ,< Img >< ImageFeature >< Img >.
After that,< Img >< ImageFeature >< Img >.
Finally   ,< Img >< ImageFeature >< Img >.
[Task Identifier ] Instruction 
\end{lstlisting}
\vspace{-2mm}
\subsection{Sound Event Timestamp Detection Module}
\vspace{-1mm}
\label{subsec:time}
\hspace{1em}In practice, sound artists need to manually determine the point in time when a sound event starts and ends and then adjust the appropriate sound effect to the correct position. This judgment is usually based on current visual information.
To simplify the process, we use the sound event timestamp detection module to detect the timestamp of the sound event in the video inspired by the CondFoleyGen \cite{duConditionalGenerationAudio2023a}.
Instead of using a hand-crafted approach to transfer sounds from the conditional audio, We use a ResNet(2+1)-D18 \cite{Tran_2018_CVPR} visual network to capture timestamps as time-conditional inputs to the LDM, trained on paired video and timestamp.

The workflow begins with feeding a sequence of silent video frames $V_f$ into the detection network.
The network then outputs a binary vector $V_{ct}$ representing predictions for each frame, derived from a fully connected layer post-pooling.
The ground truth $V_{ct}$ is obtained: Function $P$ detects audio $a$ in each frame, applying a threshold $x$ to reduce noise effects.
Sounds within 0.02s across consecutive frames are considered a single event.
This method identifies the start $x_{start}$ and end points $x_{end}$ of sound segments, as delineated in Eq.~\ref{eq:equation9}.
Based on timestamps, we construct $ T_{ct} $, as depicted in Eq.~\ref{eq:tct}, where 1 indicates the presence of sound, and 0 indicates the absence. The process equations are as follows:
\vspace{-2mm}
\begin{equation}
 x_{s t a r t}, x_{e n d}=P\left(a_{c}\right),
\label{eq:equation9}
\end{equation}
\vspace{-5mm}
\begin{equation}
T_{ct} =\begin{cases}
1 \ \ \ , if\ \ t\ \ in \ \ \ [ x_{start} ,x_{end}], \\
0 \ \ \ , else.
\end{cases}
\label{eq:tct}
\end{equation}

\begin{figure}
 \centering 
 \includegraphics[width=0.45\textwidth]{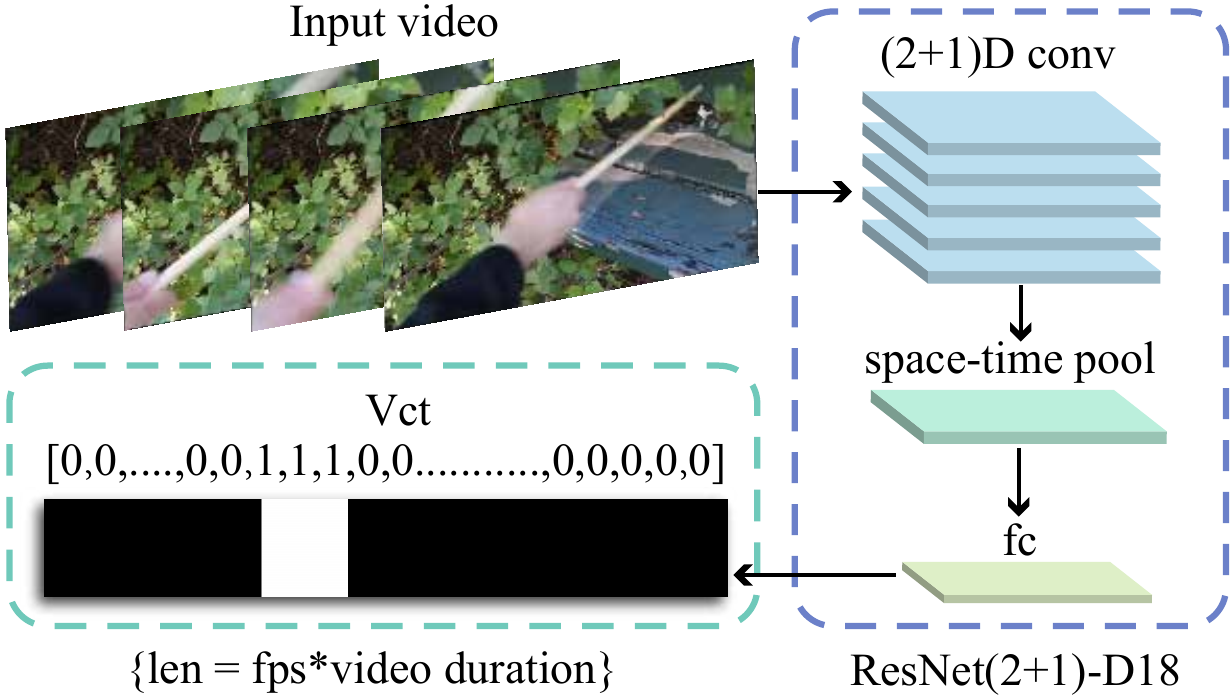}
 \caption{\textbf{ Sound Event Timestamp Detection Module.} The network analyzes the video's features to output a binary vector corresponding to the video's frame count. Within this vector, sections marked in white (value of 1) mean sound presence, and those in black (value of 0) indicate sound absence.}
 \vspace{-4mm}
 \label{fig:timestamps}
\end{figure}

Finally, $T_{ct}$ is adjusted to correspond with the video frames' duration, represented as $V_{ct}$.
We employ binary cross-entropy loss to penalize inaccuracies in time prediction.
Given that an input video may contain multiple sound events, each sound event's weights are based on its duration relative to the total sound duration in the video. 
The binary vector $T_{ct}$ serves as input for the Time-controllable Latent Diffusion Model.
The whole process and network structure are shown in Fig.~\ref{fig:timestamps}. 
\subsection{Time-controllable Latent Diffusion Model}
\label{subsec:ldm}
\hspace{1em}In our experimentation, we observed that the results generated were semantically inaccurate and temporally unsynchronized.
This issue often arose when utilizing audio-visual datasets to train end-to-end models, limiting the task's practical application.
We attribute these shortcomings to the complexity of the sound sources and the poor audio quality of the audio-visual dataset used for training.
To address this, we use text to bridge audio and video and then introduce time control in the T2A generation model.
In the visual domain, many works are based on the ControlNet \cite{AddingConditionalControl2023} architecture, which can finer control the generation of images or videos by manipulating the input image conditions of neural network blocks.
However, unlike the visual domain's inherent intuitiveness, audio features exhibit temporal continuity and are inherently more complex and abstract.
Thus, selecting appropriate audio features to guide the generation process poses a greater challenge.

In this paper, we propose a embedding called \textbf{Audio Time-condition Embedding}, denote as $A_{ct}$, which is constructed through the following procedure: 
During the training phase, the adapter extracted Mel-spectrogram features from audio waveform $a$ and then normalized it, denoted as $a_{mel} = mel(a)$.
For $a_c = max(mel(a), T_{ct})$, we then initialize $ a_c $ to match $ a_{mel} $'s dimensions but with zero values, then fill $ a_c $'s Mel channels with the maximum values based on the corresponding frames where $ T_{ct} $ equals 1.
Finally, $A_{ct}$ is derived by encoding the embedding of $a_c$ via the encoder $E_{a}$, and described as follows:
\vspace{-2mm}
\begin{equation}
\begin{aligned}
A_{ct}=E_{a}\left(\operatorname{max}\left(\operatorname{mel}\left(a_{c}\right), T_{c t}\right)\right).
\end{aligned}
\label{eq:equation10}
\end{equation}

Inspired by ControlNet \cite{AddingConditionalControl2023}, we have developed a network architecture named \textbf{Time-controllable Adapter}.
Then, we integrate the audio time-condition embedding $A_{ct}$ with the text embedding $\tau $ and target audio embedding into the neural network block.
This integration facilitates joint training for the time-controllable adapter.
$A_{ct}$ is fed solely into the adapter, while text embedding $\tau$ inputs into both Tango and the adapter.

Tango's denoising model, akin to UNet, includes encoder $F$, middle block $M$, and decoder $G$.
$F$ and $G$ have 12 corresponding blocks.
Tango's outputs of the encoder's $i-th$ block and decoder's $j-th$ block are $f_i$ and $g_j$, with $ m $ for the single middle block.
The adapter mimics Tango's encoder and middle layers as $F'$ and $M'$.
Adapter's outputs, like $f_i'$ and $m'$, are marked with ($'$).
Following this, the output of the time-controllable adapter is concatenated with the output of the corresponding decoder block during the decoding process.
For example, $f_1$ from the 1st encoder layer adds to $g_{12}$ of the decoder, making $i + j = 13$.
To achieve it, we ensure that all Tango's elements are kept frozen while modifying the input of the $i-th$ block of the decoder as:

\vspace{-2mm}
{\scriptsize
\begin{equation}
   \begin{cases}
   Concat\left( m+m^{'} ,f_{j} +zero \left( f_{j}^{'}\right)\right)    \ \ i=1, j=12.
   \\
   \\
   Concat\left(g_{i-1},f_{j}+zero\left( f_{j}^{'}\right)\right) 2 \leqslant i \leqslant 12,i+j=13.\\
   \end{cases}
 \label{eq:equation10}
\end{equation}
}
\vspace{-2mm}
\\
$zero(\cdot)$ is one zero convolutional layer, facilitating trainable and fixed neural network block connections. Its weights evolve from zero to optimized values during training.
This approach not only retains the Tango's capability for generating audio,  trained on billions of audio-text pairs, but also enables the model to comprehend the guidance provided by the time control embedding, resulting in temporally controllable outcomes. 
\vspace{-3pt}
\renewcommand{\arraystretch}{1.2} % Reduced vertical spacing
\begin{table}[!htp]
	\centering
    \setlength{\tabcolsep}{2.5mm} % Reduced space between columns
        \setlength{\abovecaptionskip}{-0.2mm}
	\caption{Distribution of CondPromptBank Categories}
	\label{mix}     
    \tiny
\begin{tabular}{lclclcl}
\hline
Category & \% & Category & \% & Category & \% \\
\hline
Household Daily & 14.11 &Transportation Vehicles & 10.42 & Impacts Crashes & 10.10 \\
Foley & 8.24 & Human Elements & 7.77 & Industrial & 6.58 \\
Weapons War & 5.83 & Cartoon Comical & 4.90 & Sports & 4.43 \\
Animals Insects & 4.04 & Instruments & 3.68 & Water Liquid & 3.27 \\
Technology & 2.70 & Horror & 2.41 & Emergency & 2.20 \\
Public Places & 1.87 & Sound Design Effects & 1.69 & Doors Windows & 1.56 \\
Fire Explosions & 1.49 & Nature Weather & 1.02 & Leisure & 0.84 \\
Multimedia & 0.47 & Bells & 0.37 &  &  \\
\hline
\end{tabular}
\label{distribution}
\end{table}
\vspace{-5mm}
\renewcommand{\arraystretch}{0.9}
\begin{table*}
  \small 
   \tabcolsep=0.24cm
 \begin{tabular}{ccccccc}
   \toprule
   Method& $CLAP-top_{general}$$\uparrow$& $CLAP-top_{unfused}$$\uparrow$& Onset Acc $\uparrow$& Onset AP$\uparrow$ & Time Acc$\uparrow$ & IoU$\uparrow$ \\
   \midrule
   CondFoleyGen(old)$^*$ \cite{duConditionalGenerationAudio2023a} & 17.3& 15.6&21.6 & \underline{67.0} & \underline{35.5} & 23.3 \\
   CondFoleyGen(new) $^*$ \cite{duConditionalGenerationAudio2023a} & \underline{16.3}& \underline{13.6}&\underline{19.2} & 68.6 & 37.4 & \underline{22.4} \\
   Ours-small& 29.6& 28.0&19.4 & 77.03 & 27.8 & 35.6 \\
   Ours-full&\textbf{36.8}& \textbf{42.8} & \textbf{27.6 }& \textbf{78.1} & \textbf{43.8} & \textbf{39.7}\\
   \bottomrule
 \end{tabular}
 \vspace{-1mm}
 \caption{\textbf{Conditional Generation Task Quantitative Results.} \textit{CLAP-top} metric evaluates a model's ability to control sound content.
It calculates the percentage of times the sound samples generated by all models ranked in the top 1 according to the CLAP ranking, divided by the total number of samples.
We used two versions of the CLAP models for evaluation: \textit{$CLAP_{general}$} and \textit{$CLAP_{unfused}$}. 
 \textit{Onset Acc} and \textit{Time Acc} are both metrics based on the number of sound occurrences, with \textit{Onset Acc} focusing on the onset count and \textit{Time Acc} on the count of time intervals. 
 We measure the average accuracy of predicting onsets within 0.1s of ground truth to assess the timing of generated onsets. \textit{IoU }is calculated by computing the intersection and union of these vectors. 
 These metrics collectively allow us to comprehensively evaluate the accuracy of the generated sounds in terms of both timing and content. 
The variant \textit{``old"} corresponds to the prior codebook, while variant \textit{``new"} matches the updated codebook. The old model was trained on 192-width spectrograms, and the new one was trained on 2s waveforms.
 ``$^*$ ''denotes the data sourced from the official code and is based on experiments conducted with our local configurations. \underline{Underline} denotes the worst performance. \textbf{Boldface} denotes the best performance. The six metrics are all measured in percentage.} 
 \vspace{-2mm}
 \label{tab:condition}
\end{table*}

\section{Experiments}
\subsection{Experiment Settings}
\begin{figure*}
  \centering
  \includegraphics[width=1\textwidth]{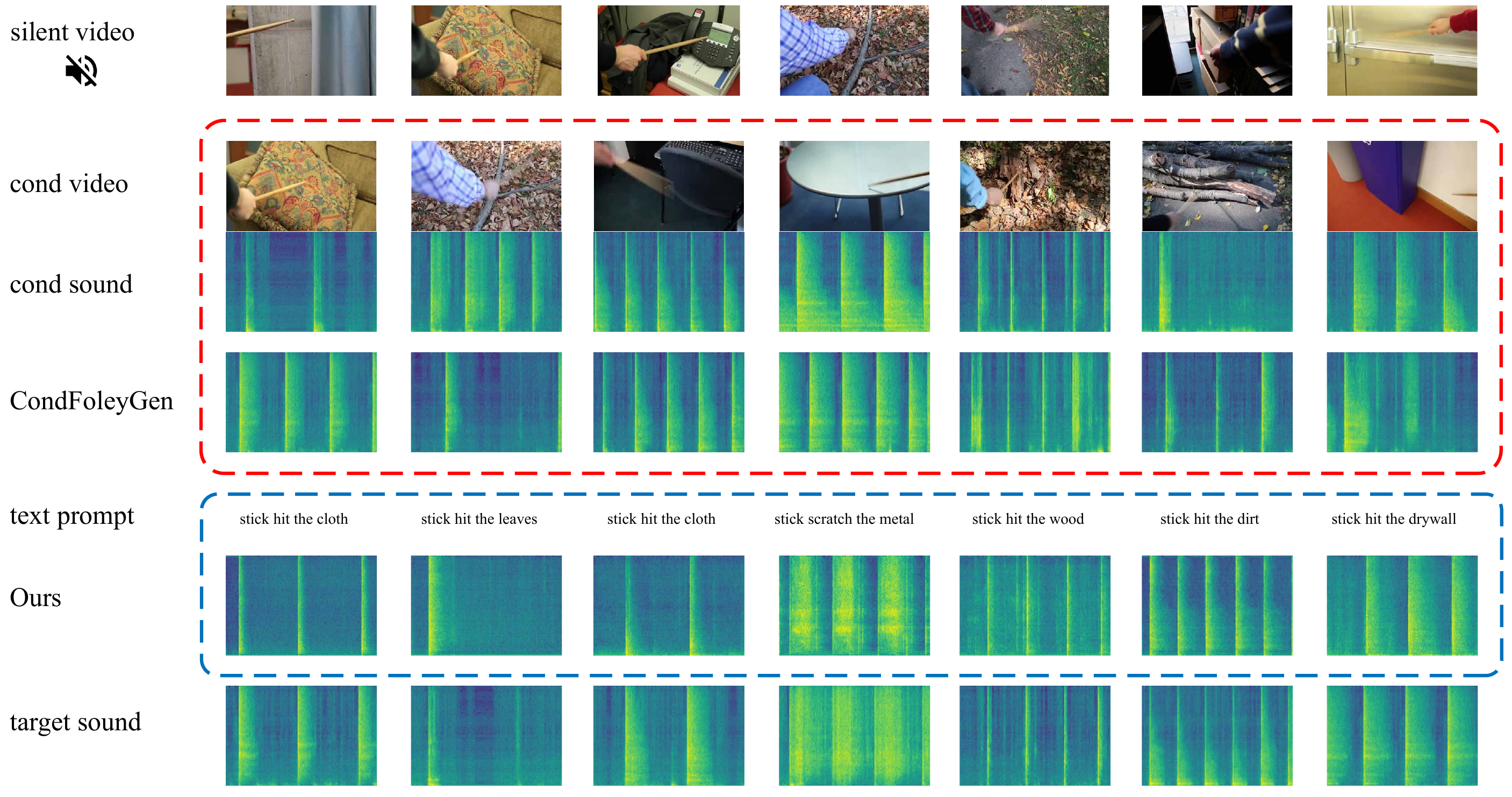}
  \caption{\textbf{Conditional Generation Task Qualitative Results.} The red dashed boxes are the conditional audio inputs and generated results for CondFoleyGen, and the blue dashed boxes are the conditional text inputs and results corresponding to SonicVisionLM.}
  \vspace{-4mm}
  \label{fig:Qualitative}
\end{figure*}

{\bf Dataset.}
Since Tango's training datasets include varied audio types like speech, sound effects, and music, it tends to produce mixed audio outputs.
Our task requires distinct handling of these audio types, ensuring each sound event is separate.
To meet our specific needs, we developed \textit{CondPromptBank}, a high-quality dataset of single sound effects crafted for training time-controlled adapters.
This dataset consists of 10,276 individual data entries, each with a sound effect, title, and start/end timestamps.
Each sound is typically 10 seconds or shorter, sourced from freely available sound effect libraries and websites.
During the collection, we focused on 23 common categories of sound effects and manually filtered out low-quality data with noise and mixed sources.
The category distribution shows Tab.~\ref{distribution}.
To enhance the textual descriptions with precise details, we further annotated the sound effect text labels with fine-grained information based on sound characteristics.
Now, each label not only identifies the audio source but also describes the associated actions in detail.
We believe the division into single audio sources is crucial, as videos provide a more concrete expression of content than text, thus requiring more precise sound representation.

\noindent{\bf Implementation details.} To train our complete model, we first train the adapter on CondPromtBank, then train the timestamps detection net on Greatest Hits \cite{Owens_gh} and Countix-AV dataset \cite{Zhang_cxav}.
We trained the adapter for approximately 200 epochs with a batch size 32 and a learning rate of $3.0 \times  10^{-5} $ using Adam \cite{kingma2017adam}.
Our model has two versions: small and full. The difference between them is the version of pre-trained parameters used for Tango.
We trained the timestamps detection net for 70 epochs with a batch size of 24 and a learning rate of $1.0 \times  10^{-5}$.
The training of the adapter takes approximately five days, and the training of the timestamps detection network requires approximately one day using one NVIDIA A6000 GPU.
\renewcommand{\arraystretch}{0.9}
\begin{table*}
  \small 
   \tabcolsep=0.50cm
  \begin{tabular}{ccccccccccccc} 
    \toprule
    & \multicolumn{6}{c}{Metric} \\ 
      \midrule
               Dataset&      \multicolumn{4}{c}{Greatest Hits}&      \multicolumn{4}{c}{Countix-AV}\\ 
Method& MKL$\downarrow$& FID $\downarrow$& IS$\uparrow$  &IoU$\uparrow$  & MKL$\downarrow$& FID  $\downarrow$& IS$\uparrow$ &IoU $\uparrow$  \\ 
 \midrule
SpecVQGAN$^*$ \cite{iashinTamingVisuallyGuided2021}& \underline{6.80}& \underline{82.4}& \underline{2.17}& 25.8& 7.39& \underline{34.1}&5.3&34.9\\  
DIFF-FOLEY$^\dag$ \cite{luoDiffFoleySynchronizedVideotoAudio2023} & 5.68& \textbf{20.0}& 3.84& \underline{22.0}& \textbf{4.9}&\textbf{15.9}&\underline{5.2}&\underline{31.3}\\
Ours-small& 6.46& 31.9&\textbf{3.88}& 36.6& \underline{10.0}& 21.7& \textbf{15.1}&37.5\\ 
Ours-full & \textbf{4.67}& 24.9& 3.26&\textbf{39.5}&9.71&19.7     &12.7&\textbf{42}\\ 
\bottomrule
  \end{tabular}
\caption{\textbf{Unconditional Sound Generation Quantitative Results.} IS assesses the quality and diversity of generated samples, 
FID measures distribution-level similarity, and MKL measures paired sample-level similarity. 
``$^*$ ''denotes the data sourced from the official code and is based on experiments conducted with our local configurations.
``$^\dag$ ''denotes that data is obtained from our adjusted official code.
\underline{Underline} denotes the worst performance. \textbf{Boldface} denotes the best performance. IoU metric is measured in percentage.}
\vspace{-4mm}
\label{tab:uncondition}
\end{table*}
\subsection{Conditional Generation Task Results}
\hspace{1em}We tested our model using the Greatest Hits \cite{Owens_gh} dataset, which has 977 videos of drumsticks interacting with different objects, lasting 11 hours. 
This dataset is divided into two types of actions and 17 types of materials. 
This detailed categorization helps check if our model can change sound types based on these details but still match the target action. 
We used the same test settings as CondFoleyGen \cite{duConditionalGenerationAudio2023a} and compared our results with theirs.
\begin{figure*}
  \centering
  \includegraphics[width=1\textwidth]{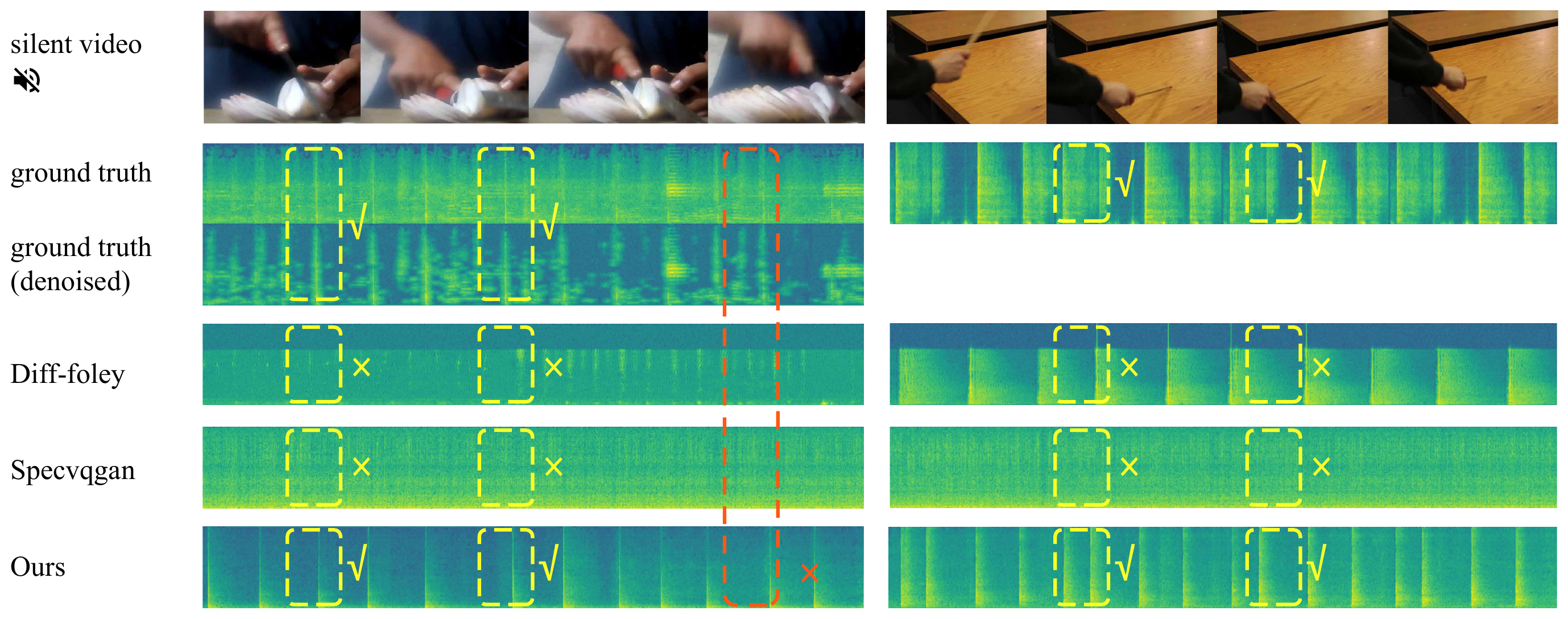}  
  \caption{\textbf{Unconditional Generation Task Qualitative Results.} The left example is from CountixAV, and the right one is from Greatest Hits. We're comparing them side by side. The dashed box highlights examples of both good and bad results we generated.}
  \vspace{-4mm}
  \label{fig:uncoditon_Qua}
\end{figure*}

\noindent{\bf Evaluation Metrics.} For the conditional generation task, we use the following five objective metrics to evaluate the performance of the model: 
\textit{CLAP-top}, \textit{Onset Acc} \cite{duConditionalGenerationAudio2023a}, \textit{Onset AP} \cite{duConditionalGenerationAudio2023a}, \textit{Time Acc}, and \textit{IoU}.

\noindent{\bf Quantitative Results.} SonicVisionLM-full demonstrates superior performance over CondFoleyGen across all metrics, as seen in Tab.~\ref{tab:condition}.
It achieves leading scores of 36.8\% and 42.8\% on \textit{CLAP-to}p versions, exceeding any CondFoleyGen model by more than 20\%.
This result suggests that it is better at matching sounds to text prompts.
Our model significantly surpasses CondFoleyGen in audio-visual synchronization, evidenced by an 11\% improvement in \textit{Onset AP} and a 16\% rise in \textit{IoU}.
Moreover, we see a 6\% enhancement in onset accuracy and an 8.3\% increase in timing precision.
These advancements illustrate our model's outstanding accuracy in sound event detection and its effectiveness in synchronizing generated sounds with the input video.
We think CondFoleyGen relies on the audio-visual synchronization module \cite{iashin2022sparse} to improve time accuracy, requiring many samples to re-ranking the sounds.
In contrast, our model gets higher synchronization during the generation process with fewer samples.

\noindent{\bf Qualitative Results.} As shown in Fig.~\ref{fig:Qualitative} (row 4, 6, 7), we compare the timestamped positional distance to the target sound.
Our results closely match the target sound, demonstrating a high degree of accuracy.
However, CondFoleyGen often produces results with the wrong number of sounds and with a large difference in the positional distance.
This indicates that our model has a higher visual-audio synchronization than CondFoleyGen.
As shown in Fig.~\ref{fig:Qualitative} (column 3, 7),
we compare the sound shape to the conditional sound.
Even though our model has not learned any timbre on the Greatest Hits dataset,
our results are still very similar to the conditional sound.
As shown in Fig.~\ref{fig:Qualitative} (column 1, 7),
The mel-spectrograms of the CondFoleyGen results do not produce sounds that are similar to, and sometimes blurred.
\subsection{Unconditional Generation}
\hspace{1em}To evaluate the task of unconditional sound generation, considering that our LDM has not been trained on audio-visual datasets, we have chosen to perform quantitative evaluation and qualitative evaluation on two datasets: Greatest Hits \cite{Owens_gh} and CountixAV \cite{Zhang_cxav}, which are zero-shot tasks for all models. 
We use two state-of-the-art V2A models as baselines:
SpecVQGAN \cite{iashinTamingVisuallyGuided2021} and DIFF-FOLEY \cite{luoDiffFoleySynchronizedVideotoAudio2023}.

\noindent{\bf Evaluation Metrics.} For \textit{objective evaluation}, we have employed three metrics as \cite{iashinTamingVisuallyGuided2021,luoDiffFoleySynchronizedVideotoAudio2023}: \textit{Inception Score (IS)} \cite{gans}, \textit{Frechet Distance (FID)} \cite{Gans_trained}, and \textit{Mean KL Divergence (MKL)} \cite{iashinTamingVisuallyGuided2021}.
For \textit{subjective evaluation}, as the \cite{duConditionalGenerationAudio2023a}, we conduct user evaluations for the three critical components: \textit{overall audio quality (OVL)}, \textit{alignment with the input video (REL)}, and \textit{time synchronization (Time-sync)}.
\renewcommand{\arraystretch}{0.9}
\begin{table}
\small 
   \tabcolsep=0.33cm
  \centering
  \begin{tabular}{cccc}
    \toprule
    Method& OVL$\uparrow$ & REL$\uparrow$ & Time-sync $\uparrow$\\
    \midrule
    SpecVQGAN$^*$ \cite{iashinTamingVisuallyGuided2021}& \underline{37}&\underline{25} &\underline{31} \\
    DIFF-FOLEY$^\dag$ \cite{luoDiffFoleySynchronizedVideotoAudio2023}& 48&64 &58   \\
    Ours& \textbf{75}&\textbf{69} & \textbf{87}  \\
    \bottomrule
  \end{tabular}
  \caption{\textbf{Subjective Results.} Following \cite{duConditionalGenerationAudio2023a}, we invited the 300 English-proficient evaluators to rate 30 randomly selected audio samples from three perspectives: \textit{OVL}, \textit{REL}, and \textit{Time-sync}. Scores were averaged on a 1-100 scale.
  ``$^*$'' denotes the data sourced from the official code and is based on experiments conducted with our local configurations.
  ``$^\dag$'' denotes data obtained from our adjusted official code.
  \underline{Underline} denotes the worst performance. \textbf{Boldface} denotes the best performance. }
   \vspace{-5mm}
  \label{tab:Subjective}
\end{table}
\noindent{\bf Quantitative Results.} In our experiments, the choice of vocoder significantly influenced the \textit{FID} metric. 
We upgraded from DIFF-FOLEY's weak Griffin-Lim vocoder to the superior MelGAN to ensure fairness.
Despite our model using 64 Mel filters compared to MelGAN's 80, our performance on the Greatest Hits dataset excels in \textit{MKL}, IS, and IoU metrics, as shown in Tab.~\ref{tab:uncondition}.
This underscores our model's precise time control and high-quality sound generation, demonstrating its exceptional capability to produce sounds that align closely with the ground truth.
In the CountixAV dataset, SonicVisionLM-small outperforms baselines by nearly 10 points in the IS metric, highlighting its exceptional sound quality.
Despite slightly lower \textit{MKL} and \textit{FID} scores compared to baselines, we do not view this as a drawback. 
Unlike the Greatest Hits dataset, which was recorded in high quality with specialized recording equipment, the CountixAV dataset is sourced from diverse YouTube videos. It often includes sounds marred by low-quality background noise or mixed sound sources, especially human vocals.
\textit{MKL} and \textit{FID} metrics emphasize similarity to the ground truth. 
Since DIFF-FOLEY was trained on similar audio-visual data, its tendency to mix multiple sounds in response to complex visual information explains its advantage in these metrics.
Our model, targeting sound events from specific actions and filtering out irrelevant auditory information, diverges from the ground truth, a deliberate choice to enhance sound event relevance.
The \textit{IoU} metrics show that our model makes sound at the correct times and remains silent otherwise.
To further argue our point, we conducted a subjective assessment of three aspects of \textit{OVL}, \textit{REL}, and \textit{Time-sync}.
As shown in Tab.~\ref{tab:Subjective}, our model performed significantly better than the baseline models in all metrics.
\\
{\bf Qualitative Results.}
For clearer time point comparison, we denoised the data from the CountixAV.
Fig.~\ref{fig:uncoditon_Qua} reveals our generated sounds are more precise than the baseline models.
Furthermore, our model can capture finer-grained visual transformations, capturing the signal despite sudden changes in the frequency of movement and producing the correct sound.
In contrast, SpecVQGAN results are full of noise.
DIFF-FOLEY can sometimes generate clear sounds, but it is not aligned with the visual-sound synchronization.
\renewcommand{\arraystretch}{0.9}
\begin{table}
\small 
   \tabcolsep=0.35cm
  \centering
  \begin{tabular}{ccccc}
    \toprule
    Method& FD$\downarrow$  & FAD$\downarrow$  & KL$\downarrow$  & IoU$\uparrow$ \\
    \midrule
    tango-small$^*$ \cite{ghosalTexttoAudioGenerationUsing2023}& \underline{44.9}&\underline{11.5}&\underline{3.47}&41.3\\
    Ours-small&26.2&\textbf{2.05}&\textbf{2.30}&\textbf{71.6}\\
    \midrule
    tango-full$^*$ \cite{ghosalTexttoAudioGenerationUsing2023}&44.3&10.87&3.14&\underline{39.3}\\
    Ours-full&\textbf{25.4}&3.19&2.32&65.7\\
    \bottomrule
  \end{tabular}
  \caption{\textbf{Time Condition Ablation Results.}
  ``$^*$ ''denotes the data sourced from the official code and is based on experiments conducted with our local configurations. \underline{Underline} denotes the worst performance. \textbf{Boldface} denotes the best performance. IoU metric is measured in percentage.
  }
  \vspace{-6mm}
  \label{tab:Ablation}
\end{table}
\begin{figure}
  \centering
  \includegraphics[width=0.48\textwidth]{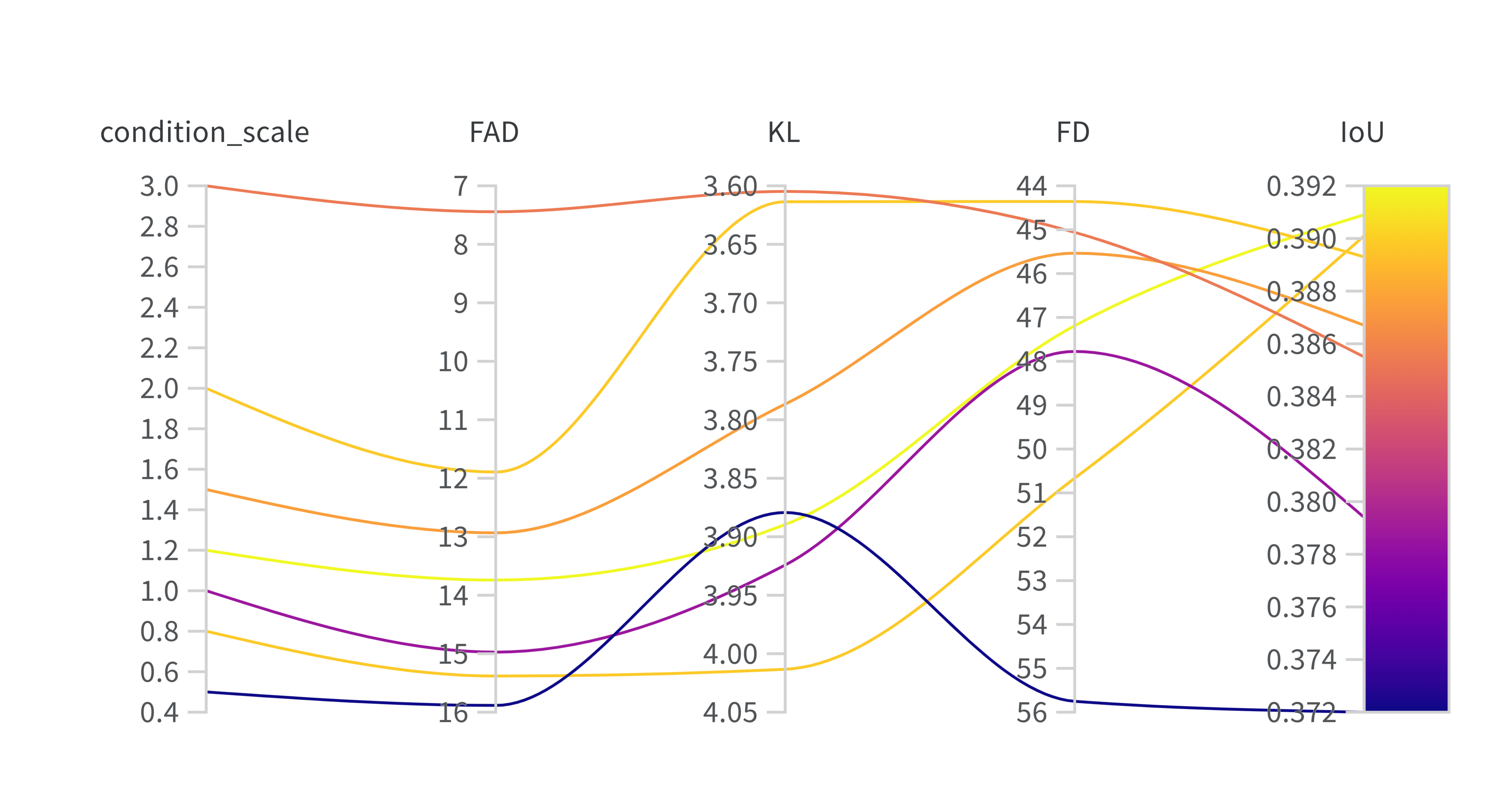}
  \setlength{\abovecaptionskip}{-6mm}
  \caption{\textbf{Conditioning-scale Ablation Study.} In the chart, the higher the indicators are, the better they are. The brightness of the colour also represents a good or bad performance.}
  \vspace{-5mm}
  \label{fig:control_scale}
\end{figure}
\begin{figure}
  \centering
  \includegraphics[width=0.48\textwidth]{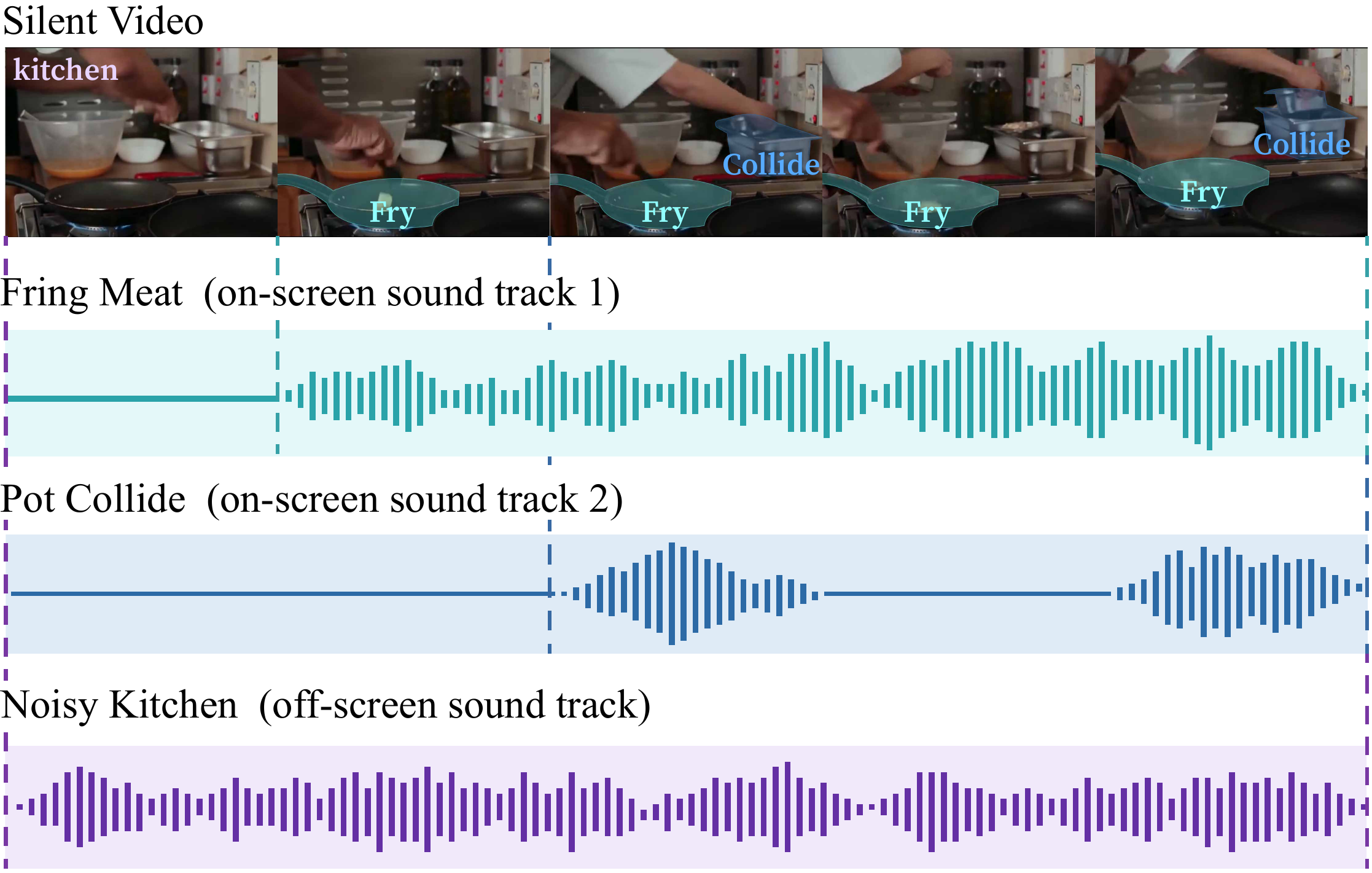}  
  \caption{\textbf{Multi-soundtracks generation example.} In the video, the blue and green masks show different soundtracks. The timing of on-screen sounds (corresponding to visible actions) aligns accurately with their occurrence in the frames. Conversely, off-screen sounds (not visible in the video) lack precise timing.}
  \vspace{-6mm}
  \label{fig:multi}
\end{figure}
\subsection{Ablation Study}
{\bf Time condition.}
\textit{Evaluation Metrics.}
 We employed objective metrics from Tango \cite{ghosalTexttoAudioGenerationUsing2023}, such as the \textit{Frechet Audio Distance (FAD)} \cite{FAD} and \textit{Frechet Distance(FD)} \cite{liuAudioLDMTexttoAudioGeneration2023} to measure the distribution difference between generated audio and ground truth without the need for any reference audio samples. 
\textit{KL divergence} \cite{diffsound,kreukAudioGen} assesses their similarity based on the broad concepts present in the original and the generated audio signals.
\\
\textit{Result.}
we compared our model with the Tango \cite{ghosalTexttoAudioGenerationUsing2023}.
As shown in Tab.~\ref{tab:Ablation}, both in the small and full versions,
our model outperforms Tango in terms of IoU, which validates the effectiveness of the time-controllable adapter.\\
{\bf Conditioning-scale.}
As seen in the Fig.~\ref{fig:control_scale}.
The sound quality improves in the range from 0.6 to 2.0 on the conditioning scale as it lines up better with the IoU. But at 3.0, control goes down a lot. The best mix of good sound quality and steadiness is found at a conditioning scale equal to 2.0.
\subsection{Multi-soundtracks generation}
\hspace{1em}As shown in Fig.~\ref{fig:multi}, SonicVisionLM works like a sound designer in traditional video post-production. 
First, it looks at the visual information to find the key sounds needed. 
For example, it picks out the hissing of cooking meat and the clinking of pots. 
Then, if the story needs it, the sound designer adds off-screen sounds. 
This might include extra kitchen sounds like distant chopping, simmering pots, or the buzz of kitchen machines, making it feel more natural. 
Lastly, SonicVisionLM puts these audio tracks in the right place on the video timeline. 
This ensures the sounds match perfectly with what is happening in the video, creating a smooth audio-visual experience.
More detailed results can be found in the Supplementary Materials.

\section{Conclusion}
\hspace{1em}In this paper, we propose SonicVisionLM, which utilizes the capabilities of powerful vision-language models (VLMs). When provided with a silent video, SonicVisionLM first identifies events within the video using a VLM to suggest possible sounds that match the video content. 
SonicVisionLM demonstrates outstanding performance in both conditional and unconditional generation tasks. 
Further, we tested its efficacy in post-production, focusing on automatic recognition of on-screen sounds and personalized editing for off-screen sounds. Extensive experiments show the superior performance of our method.
\\
{\bf Limitations.} While SonicVisionLM has achieved adequate results, refinements in the visual understanding and timestamp detection parts are still required.
Regarding various multimedia contexts, expanding the diversity and range of audio generation control are still the key demands of applications.
Thus, the model should enrich the audio-visual experience and broaden applicability in future attempts.
{
    \small
    \bibliographystyle{ieeenat_fullname}
    \bibliography{main}
    
}

\clearpage
\twocolumn[{%
\renewcommand\twocolumn[1][]{#1}%
\maketitlesupplementary
\begin{center}
    \vspace{-3mm}
    \includegraphics[width=0.9\textwidth]{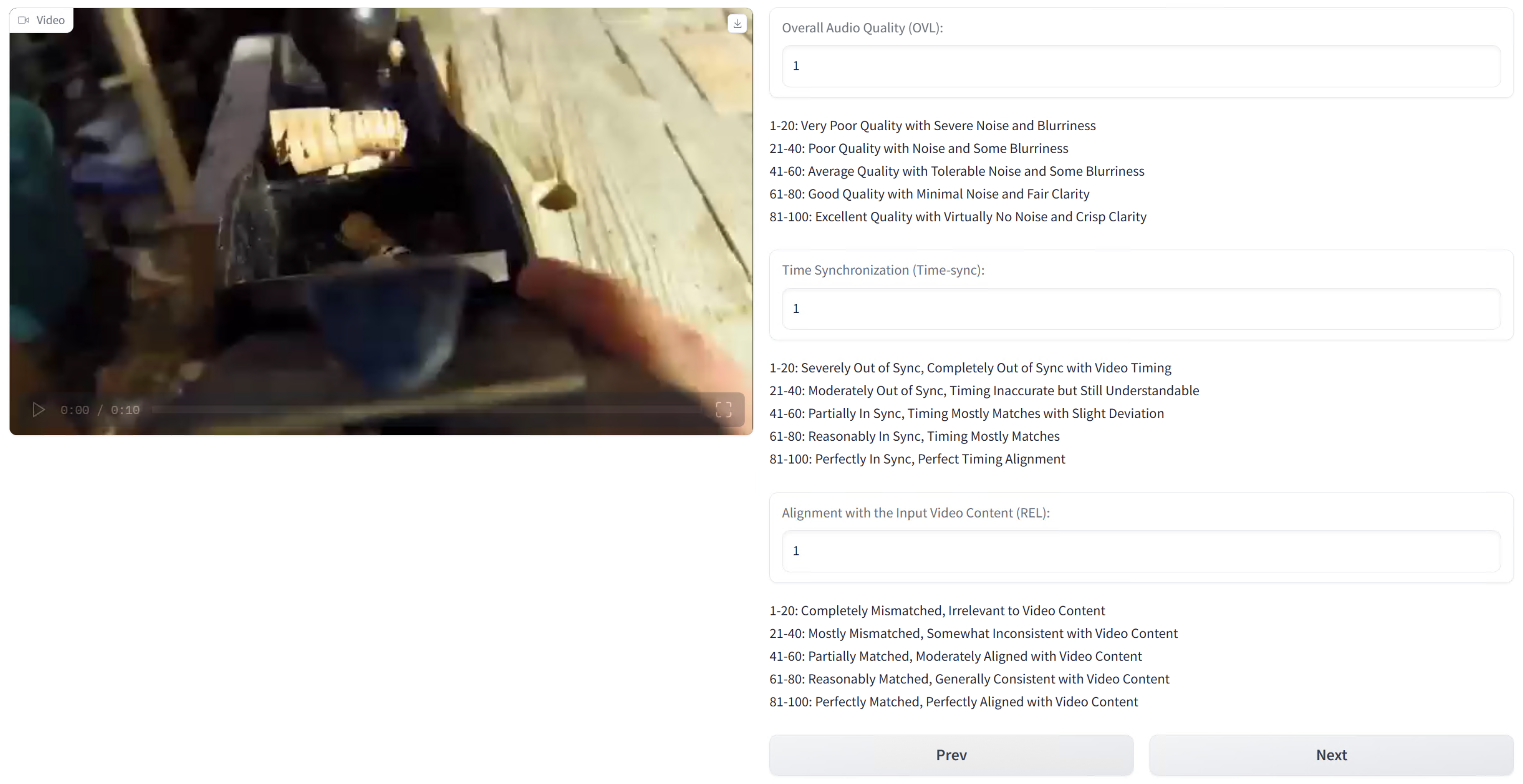}
    \captionof{figure}{The evaluation page for the unconditional generation task. We display a screenshot of the primary test interface that participants will encounter. Each participant is required to input three scores for the presented content. Upon clicking the 'Next' button, they will be directed to the subsequent video in the test sequence.}
    \label{fig:user_study}
    \vspace{3mm}
\end{center}%
}]

\renewcommand{\thesection}{A.\arabic{section}}
\setcounter{section}{0}
\section{Subjective Results Details}
We provide the screenshot of the main evaluation page the participant will see during the test in Fig.\ref{fig:user_study}.
\section{Timestamp Detection Module Precision Experiment}
Our timestamp detection model shows promising performance, with an Average Precision(AP) of 0.80 and an accuracy of 0.72 on the Greatest Hits test set, and even higher results on the validation set with an AP of 0.92 and accuracy of 0.82. However, its performance on the CountixAV test set is comparatively lower, achieving an AP of 0.52 and an accuracy of 0.52. This discrepancy likely stems from the complexity of sound sources in our training dataset, leading to potential inaccuracies in ground truth. Such labelling challenges can adversely affect recognition accuracy, particularly in scenarios involving non-static footage.
\begin{figure}
  \centering
  \includegraphics[width=0.5\textwidth]{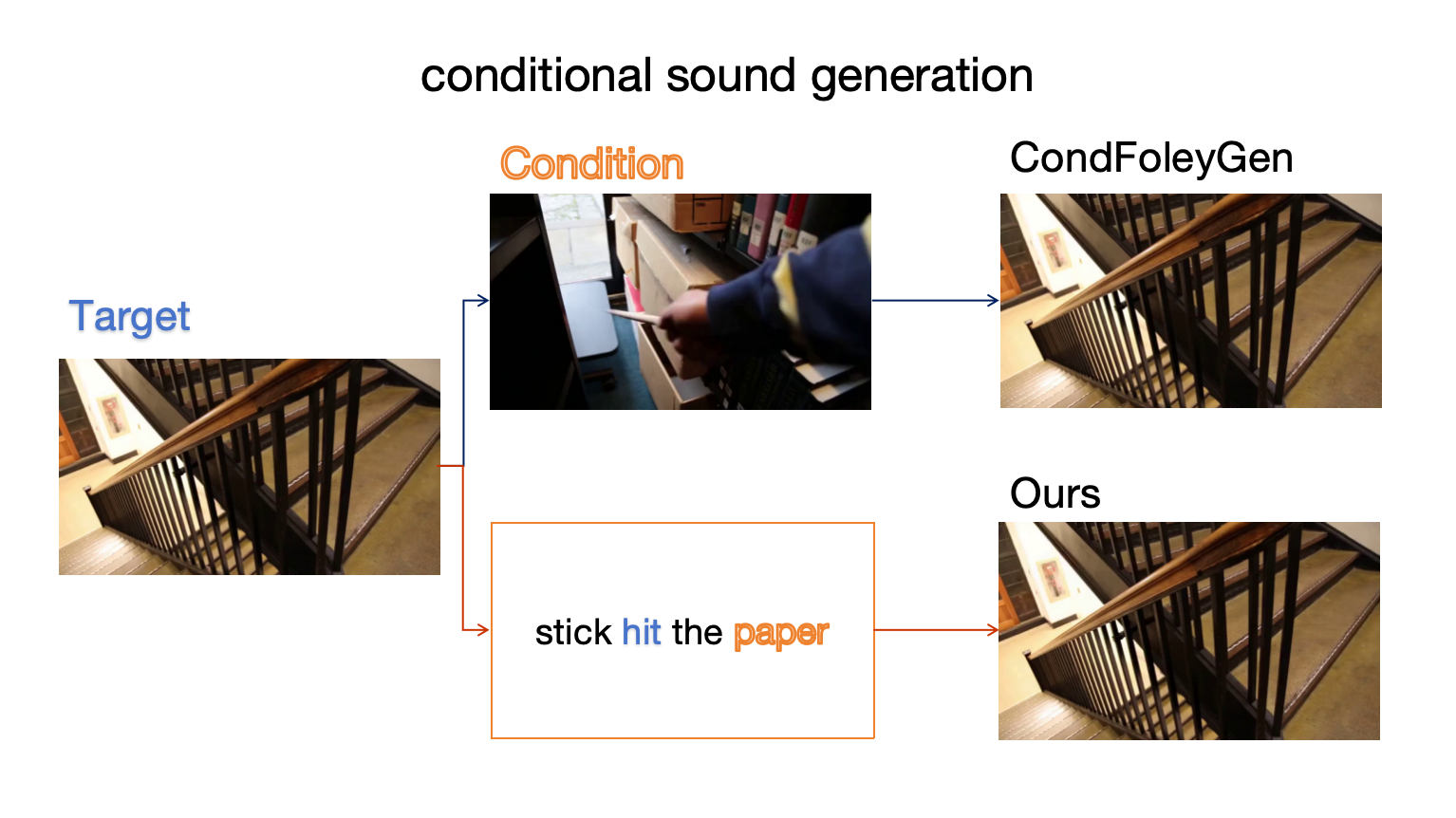}  
  \caption{screenshot of the conditional sound generation task section in the demo video.}
    \vspace{-3mm}
  \label{fig:compare1}
\end{figure}
\section{Additional Results}
 \begin{figure}
  \centering
  \includegraphics[width=0.45\textwidth]{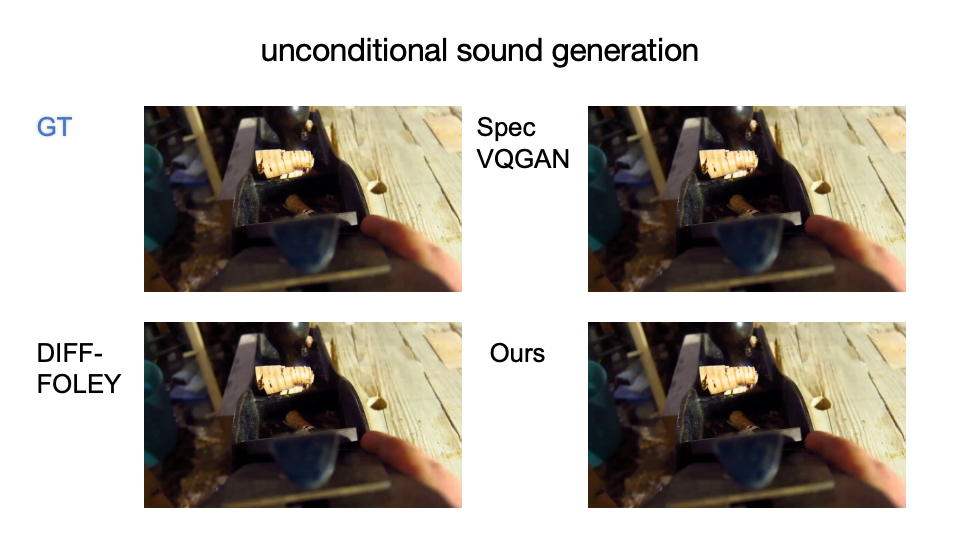}  
    \vspace{-3mm}
  \caption{screenshot of the unconditional sound generation task section in the demo video.}
  \label{fig:compare2}
\end{figure}
 \begin{figure}
  \centering
  \includegraphics[width=0.45\textwidth]{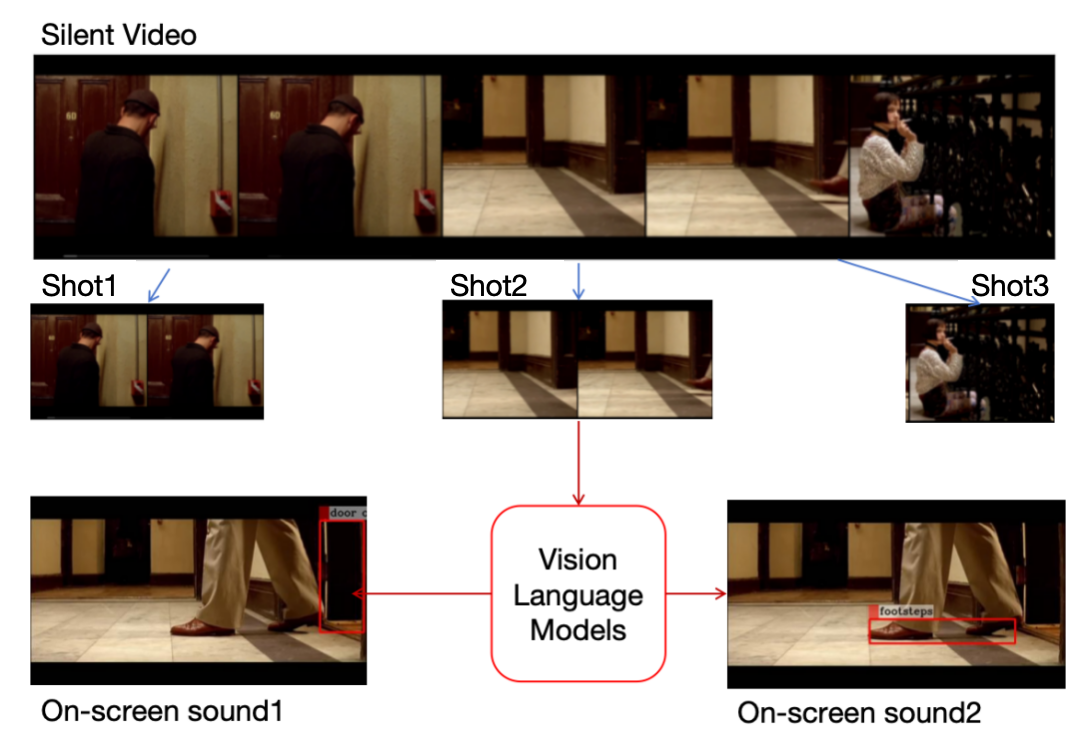}  
  \vspace{-4mm}
  \caption{The process of video-to-text and text-based interaction components in multi-track generation tasks.}
  \label{fig:muti_soundtracks}
\end{figure}
{\bf Conditional Generation Results.} As shown in Fig.\ref{fig:compare1}, the left column represents the target video, the middle column showcases the control conditions for CondFoleyGen and our model, and the right column displays the generated results. We provide 6 examples for the previously mentioned conditional generation task, and the corresponding audio outcomes can be observed in the demo video.
 
\noindent{\bf Unconditional Generation Results.} As shown in Fig.\ref{fig:compare2}, we compare a comparison of the results generated by GT, SpecVQGAN, DIFF-FOLEY, and our model. We provide 6 examples from CountixAV, and the audio samples are available in the demo video.

\noindent{\bf Multi-soundtracks Generation Results.} As shown in Fig.\ref{fig:muti_soundtracks}, we segment the video into different shots first. Shots with clear actions are directly processed by the VLMs to obtain corresponding sound effect descriptions and their spatial positioning within the video. These descriptions are used to generate on-screen sound. 
Other shots accept user editing and are then fed into the LDM to produce off-screen sounds. We provide a simple and a complex case respectively in the demo video.

\end{document}